\documentstyle[galley]{mn}
\twocolumn
\input{epsf}
\def\Real{{\rm I\mathchoice{\kern-0.70mm}{\kern-0.70mm}{\kern-0.65mm}%
  {\kern-0.50mm}R}}

\font \bolditalics = cmmib10
\def\bx#1{\leavevmode\thinspace\hbox{\vrule\vtop{\vbox{\hrule\kern1pt
        \hbox{\vphantom{\tt/}\thinspace{\bf#1}\thinspace}}
      \kern1pt\hrule}\vrule}\thinspace}

\def \vc #1{{\textfont1=\bolditalics \hbox{$\bf#1$}}}

\def\kg{{\bf k}}

\def\rg{{\bf r}}

\def\ggr{{\bf g}}

\def\thetag{{\vc \theta}}

\def\gammag{{\vc \gamma}}
\def\omegag{{\vc \omega}}
\def\Gammag{{\vc \Gamma}}
\def\epsilong{{\vc \epsilon}}
\def\Sigmag{{\vc \Sigma}}
\def\deltag{{\vc \delta}}

\def\Vg{{\vc V}}
\def\Mg{{\vc M}}

\def\Pc{{\cal P}}

\def\Ac{{\cal A}}

\def\Nc{{\cal N}}

\def\Dc{{\cal D}}

\def\Sc{{\cal S}}

\def\be{\begin{equation}}
\def\ee{\end{equation}}
\def\ba{\begin{eqnarray}}
\def\ea{\end{eqnarray}}

\def\d{{\rm d}}

\title{Noise properties of gravitational lens mass reconstruction}

\author[Van Waerbeke]{Ludovic Van Waerbeke \\
Canadian Institute for Theoretical Astrophysics, 60 St George Str.,
Toronto, M5S 3H8, Canada\\
\\}

\pagerange{000,000}

\begin{document}

\maketitle
\begin{abstract} 
Gravitational lensing is potentially able to observe mass-selected halos,
and to measure the projected cluster mass function. An optimal
mass-selection requires a quantitative understanding
of the noise bahavior in mass maps. This paper is an analysis of the noise
properties in mass maps reconstructed using a maximum likelihood method.

The first part of this work is the derivation of the noise power spectrum and
the mass error bars as a straightforward extension of the Kaiser \& Squires (1993)
algorithm to the case of a correlated noise. A very good agreement is found
between these calculations and the noise properties observed in
the mass reconstructions limited to non-critical cluster of
galaxies. It demonstrates that Kaiser \& Squires (1993) and maximum likelihood
methods have similar noise properties and that the weak lensing approximation
is valid for describing it.

In a second stage I show that the statistic of peaks in the noise
follows accurately the peak statistics of a two-dimensional Gaussian
random field (using the BBKS technics) if the smoothing aperture contains
enough galaxies. This analysis provides a full procedure to derive the significance
of any convergence peak as a function of its amplitude and its profile.

It is demonstrated that, to a very good approximation, a mass map is 
the sum of the lensing signal plus a 2D gaussian random noise, which 
means that a detailled quantitative analysis of the structures in
mass maps can be done. A straightforward application is the
measurement of the projected mass function in wide field lensing surveys,
down to small mass over-densities which are individually undetectable.

\end{abstract}

\section{Introduction}

Non-parametric mass reconstruction from gravitational shear
is now a very well understood problem, and was widely
refined over the last few years. It is no longer a technical issue
to reconstruct clusters (see Squires \& Kaiser 1996 for
a detailed review) as well as large scale
structures (Van Waerbeke et al. 1999). Although the computational
speed of mass reconstruction might be too low for large
scale surveys (Lombardi \& Bertin 1999), this is not a major concern yet according
to the actual size of the lensing surveys.

Naturally a next step in gravitational
lensing is the search for a better use of the information present in the mass maps.
Assuming that all the problems concerning the residual of Point Spread Function
corrections are or will be solved, there is useful information hidden in the noise
of mass maps that can be extracted. An important application
of lensing is the possibility
to built mass-selected samples (Reblinsky \& Bartelmann, 1999) as a probe of
cosmology (Oukbir \& Blanchard 1997, Reblinsky et al. 1999). In
particular a joint analysis of clusters of galaxies from
weak lensing, Sunyaev-Zeldovich, X-rays and
galaxy catalogues would be valuable. A related
issue is the possible existence of {\it dark lenses} (Hattori 1997),
and already, the detection of a mass clump not obviously associated
with light (Erben et al. 1999)
raised the question of quantifying accurately the significance of halo detections.

The understanding of the noise properties in mass maps is therefore necessary
for anyone who wants to measure the mass distribution from gravitational lensing
in order to constrain cosmological models.
This point was understood by Schneider (1996) who constructed the so-called
S-statistic, aimed
to quantify the significance of mass peaks. Using bootstrapping of the data, he
derives the frequency of observed mass peaks. The S-statistic is similar to the
the signal-to-noise derived in Kaiser \& Squires (1993) (and confirmed later
in numerical simulations by Wilson et al. 1996) but it has the interesting
feature that works directly with the ellipticity of the galaxies (no mass
reconstruction is required), and the smoothing filter can be optimized in order
to increase the signal-to-noise.

So far, the noise analysis and the statistical significance of mass
reconstructions have not been studied as much as the
mass reconstruction methods themselves. We have only some global
clues of the behavior of the noise, which above all, seem to depend
on the mass reconstruction method (Squires \& Kaiser 1996). Seitz \& Schneider
(1996) identified several characteristics of the noise in their numerical
simulations: due to the smoothing, the power spectrum is not flat and
driven by the intrinsic ellipticity of the galaxies, their number
density and the smoothing window scale. Lombardi \& Bertin
(1998a, 1998b) gave the first analytical estimate of the variance of the
total mass of a reconstructed cluster, taking into account the noise correlation
induced by the smoothing window.
They have found differences with the numerical results of Seitz \& Schneider
(1996) that they attributed to {\it technical} differences.
They worked with shear line-integral reconstruction methods, and
found that the noise is minimized when curl-free kernels
are used.
As discussed in Seitz \& Schneider (1996) the curl-free kernels have
a noise amplitude similar to the Kaiser \& Squires (1993) (hereafter KS)
reconstruction, but the line-integral reconstruction is a non-local
process, which complicates the noise analysis (Lombardi \& Bertin
(1998a, 1998b)). On the other hand, the Maximum Likelihood (hereafter ML)
algorithm (Bartelmann et al. 1996) performs a local reconstruction, but its noise
properties can hardly be studied analytically. In fact, the KS scheme seems to be
the easiest approach to an analytical description of the noise.

It has been demonstrated in the case of lensing by large scale structures
(see the simulations performed in Van Waerbeke et al. 1999) that the noise in ML
reconstructions is accurately
described by the weak lensing approximation (and therefore by the KS equations).
Continuing this work, this paper is a study of the noise properties of
ML methods in the case of non-critical clusters of galaxies with
correlated noise, when the smoothing is taken into account. 

This paper shows that the noise in ML reconstruction agrees very well
with the analytical prediction
made from the weak lensing approximation. It demonstrates that the mass variance
derived in Lombardi \& Bertin (1998b) is valid, and, to a larger extent,
that the noise is uncorrelated with the signal even for non-critical clusters of
galaxies. This is also consistent with the empirical analysis done in
Seitz \& Schneider (1996).
Assuming that the smoothed intrinsic ellipticities are described by
the statistics of a two-dimensional Gaussian random field, the BBKS
technique
(Bardeen et al. 1986) is used in order to quantify the significance
of the over-densities in ML mass reconstructions. This analysis
provides a quantitative way to identify the structures in mass maps
and to extract information from the noise: this result is verified
on numerical simulations, which shows that 
to a very good approximation a mass map is the sum of a completely known
Gaussian field plus the lensing signal.

Section 2 gives some general considerations about lensing theory used in this work.
The noise model is developed in Section 3, where each noise contribution is
identified. An estimator for the variance of the mass in a given area
is calculated. In Section 4, the noise model and numerical
simulations of cluster
lensing are compared. In Section 5 the significance
level of a reconstructed mass halo based on a 2D Gaussian field
statistic is derived, which is then checked against numerical simulations.

\section{Generalities}

We consider a lens of surface mass density $\Sigma(\thetag)$ located
at a redshift
$z_l$ along the line of sight $\thetag$, and a single source plane is put
at a redshift $z_s$. At first approximation, the Universe is equivalent to
a single lens, so there is no loss of generality here to consider a single lens.
The projected
gravitational potential $\Phi$ and the surface mass density
of the lens are given by the 2-dimensional Poisson equation

\begin{equation}
\Delta\Phi(\thetag)=4\pi G\Sigma(\thetag).
\end{equation}
The critical density (for which the lens produces multiple images) is given by

\begin{equation}
\Sigma_{crit}={c^2\over 4\pi G} {\Dc_{os}\over \Dc_{ol}\Dc_{ls}},
\label{sigma_crit}
\end{equation}
where $\Dc_{ij}$ is the angular diameter distance between the redshifts
$z_i$ and $z_j$. For convenience, the lensing potential is defined
as $\phi=\Phi/(2\pi G \Sigma_{crit})$. The ratio

\begin{equation}
\kappa(\thetag)={\Sigma(\thetag)\over \Sigma_{crit}},
\label{kappa_def}
\end{equation}
is the convergence of the lens, which describes the isotropic
deformation of a light beam coming from the sources.
The light beam is also sheared, which modifies the shape of the
source galaxies. The shape of a
galaxy is described by the second moments of its centered surface
brightness profile

\begin{equation}
Q_{ij}={\int \d\thetag~\theta_i\theta_j~{\cal S}(\thetag)\big/
\int \d\thetag{\cal S}(\thetag)}.
\label{moments_def}
\end{equation}
The source and image shape matrices are related by the equation

\begin{equation}
Q^{(I)}_{ij}=\Ac Q^{(S)}_{ij}\Ac,
\label{mapping}
\end{equation}
where $\Ac$ is the magnification matrix which depends on the convergence
$\kappa$ and the shear $\gammag$ of the lens,

\begin{equation}
{\cal A}=\pmatrix{1-\kappa-\gamma_1& -\gamma_2 \cr -\gamma_2 & 1-\kappa+\gamma_1
}.
\end{equation}
$\kappa$ and $\gammag$ are fully determined by the second derivatives
of the lensing potential via

\begin{eqnarray}
\kappa=\Delta \phi/2; \ \
\gamma_1=\left(\phi_{11}-\phi_{22}\right)/2;\ \
\gamma_2=\phi_{12}.
\label{lens_def}
\end{eqnarray}
The ellipticity is defined in the complex plane as

\begin{equation}
\epsilong={1-r\over 1+r}{\rm e}^{2{\rm i}\vartheta},
\label{elli_distrib}
\end{equation}
where $r$ is the axis ratio of a galaxy and $\vartheta$ the orientation.
Eq.(\ref{mapping}) written in terms of source and image ellipticities is

\begin{equation}
\epsilong^{(I)}={\epsilong^{(S)} +\ggr\over 1+\ggr^\star\epsilong^{(s)}},
\label{elli_single}
\end{equation}
where $\ggr=\gammag/(1-\kappa)$ is the complex reduced shear.
The observable is $\deltag=2\ggr/(1+|\ggr|^2)$, but
this paper is restricted to non-critical lenses, for
which $|\deltag|<1$, so that $\ggr$ is an observable as well. In that
case $\ggr$ can be immediately read off from the galaxy ellipticities; as shown
by Schramm \& Kayser (1995) and Schneider \& Seitz (1995) the mean observed
ellipticities of the galaxies is an unbiased estimate of the reduced shear:

\begin{equation}
\langle \epsilong^{(I)} \rangle=\ggr.
\end{equation}
The mass reconstructed from shear measurement alone is not fully constrained
because any transformation of the
lensing potential of the form

\begin{equation}
\phi \rightarrow \lambda\phi+{(1-\lambda)\over 2}|\thetag|^2
\label{transfo}
\end{equation}
leaves the reduced shear $\ggr$ unchanged. This is the so-called mass-sheet degeneracy:
the mass is determined up to an additive constant and an appropriate
scaling (Eq.(\ref{transfo}) implies $\kappa \rightarrow 1-\lambda+\lambda\kappa$).

\section{Noise correlation and mass error bar}

\subsection{Weak lensing approximation}

As already mentioned, the weak lensing approximation
linearises the lensing equation and dramatically simplifies
the noise analysis.
The weak lensing limit of Eq.(\ref{elli_single}) is

\begin{equation}
\epsilong^{(I)}\simeq\gammag+\epsilong^{(S)}.
\label{weaklenslimit}
\end{equation}
Thus $\langle \epsilong^{(I)}\rangle=\gammag$ becomes an unbiased estimate of
the shear. In the Fourier space
Eq.(\ref{lens_def}) takes the form (Kaiser 1998):

\begin{equation}
\tilde\kappa(\kg)=c_\alpha(k)\tilde\gamma_\alpha(\kg),
\label{lens_eq}
\end{equation}
where there is an implicit summation over $\alpha=(1,2)$.
$c_\alpha=(\cos (2
\varphi)),\sin(2\varphi))$, and $\kg=k(\cos\varphi,\sin\varphi)$.
An unbiased (but extremely noisy) estimate of the convergence $\kappa$ is
$\kappa_n(\thetag)=\kappa(\thetag)+n(\thetag)$, where $n(\thetag)$ is the
noise part, which depends on the intrinsic ellipticities of the galaxies
and their spatial distribution. The Fourier transform of $\kappa_n$ is

\begin{equation}
\tilde\kappa_n(\kg)=c_\alpha\tilde\epsilon^{(I)}_\alpha(\kg)=
\tilde\kappa(\kg)+c_\alpha\tilde\epsilon^{(S)}_\alpha(\kg).
\label{wl_eq}
\end{equation}
In order to avoid the infinite variance problem (caused by the discreteness
distribution of the galaxies), and to have a mass map
with a reasonable amplitude of noise, the convergence is reconstructed
from the smoothed galaxy ellipticities, so
smoothed fields are used in the following. Uppercase
letters refer to smoothed fields while lowercase letters refer to
unsmoothed fields (e.g. $K$ for the smoothed $\kappa$, $\Gammag$ for
$\gammag$ and $\Sigmag$ for $\epsilong$).
The smoothing function is denoted by $W(|\thetag|)$, the total number of galaxies in
the field by $N_g$, and the total field area by $\Sc$. Galaxies are at positions
$\thetag_i$. The reconstructed mass map is now the smoothed convergence
with an unbiased estimate given by

\begin{equation}
K(\thetag)={\Sc\over N_g}{\displaystyle\sum_{i=1}^{N_g}} W(|\thetag-\thetag_i|)\kappa(\thetag_i)
\label{itissmooth}
\end{equation}
The number density of galaxies is $n_g=N_g/\Sc$.
It is worth noting that Eq.(\ref{lens_eq}) implies that the reconstructed
convergence of a smoothed shear field is identical to the initial
smoothed convergence. In other words, smoothing and reconstruction are
commutative processes in the frame of the weak lensing approximation.

\subsection{Noise correlation function}

The reconstructed convergence
$K_N(\thetag)$ is the sum of the true convergence
$K(\thetag)$ and a correlated noise $N(\thetag)$.
If $\Sigmag^{(I)}$ and $\Sigmag^{(S)}$ are the smoothed
$\epsilong^{(I)}$  and $\epsilong^{(S)}$ then

\begin{equation}
K_N(\thetag)=K(\thetag)+N(\thetag)=\int{\rm d}\kg~e^{-i\kg\cdot\thetag}
c_\alpha(k)\tilde\Sigma^{(I)}_\alpha(\kg).
\label{kappanoise}
\end{equation}
If the smoothing length is small enough (arcmin scale) the signal is
approximately constant within the aperture. This is certainly not
true for cluster mass profiles, but it should still give a reasonable
description of the noise. It implies that the sum over the galaxy ellipticities
within the aperture is equal to the smoothed shear plus the intrinsic
ellipticity contribution (which is Eq.(\ref{weaklenslimit}) for the
smoothed fields):

\begin{equation}
\Sigmag^{(I)}(\thetag)\simeq\Gammag(\thetag)+{1\over n_g}{\displaystyle 
\sum_{i=1}^{N_g}}W(\thetag-\thetag_i)\epsilong^{(S)}(\thetag_i).
\label{approx_shear}
\end{equation}
Inserting this into Eq.(\ref{kappanoise}) gives the analytic
expression of the noise term:

\begin{equation}
K_N(\thetag)-K(\thetag)\simeq {1\over n_g}{\displaystyle\sum_{i=1}^{N_g}}
\int{\rm d}\kg~\tilde W(\kg) e^{-i\kg\cdot(\thetag_i-\thetag)}c_\alpha(k)
\epsilon_\alpha(\thetag_i),
\label{noise_term}
\end{equation}
where $\tilde W(\kg)$ is the Fourier transform of the window function.
The ensemble average of the square of Eq.(\ref{noise_term}) is the noise
variance $\langle N^2(\thetag)\rangle$. It is calculated by
averaging over the intrinsic ellipticities
and the galaxy positions. Let $\rm A$ be the operator which averages over the
intrinsic ellipticities and $\rm P$ the operator which averages over the
galaxy positions. The former is defined as

\begin{equation}
\rm A\left[\epsilon_i^{(S)}(\thetag)\epsilon_j^{(S)}(\thetag')\right]=
{\sigma_\epsilon^2\over 2} \delta_{ij}^K \delta_D(\thetag-\thetag').
\label{A_operator}
\end{equation}
$\delta_{ij}^K$ is the Kronecker symbol, and $\sigma_\epsilon$ the 
intrinsic ellipticity dispersion. This is a consequence of the fact that
the galaxy ellipticities are assumed to be spatially uncorrelated. The galaxy
position average operator
$\rm P$ depends on the angular clustering of the galaxies
within the aperture (Peebles 1980):

\begin{equation}
{\rm P}=\int{{\rm d}^2\thetag_1{\rm d}^2\thetag_2\over \Sc^2}\left(1+\omega(
|\thetag_2-\thetag_1|)+...\right)
\label{P_operator}
\end{equation}
where $\Sc$ is total field area, and $\omega(|\thetag_2-\thetag_1|)$ is the
angular two point correlation function of the source galaxies.
This clustering term is generally ignored, and accounts for possible
increase of the variance
of $N(\thetag)$ in the case of non-uniform sampling of the source
galaxies within the aperture.
Its contribution can be potentially important if the source galaxies
belongs to a thin redshift slice (as the angular
clustering approaches the spatial clustering). Connolly
et al. (1998) measured on the Hubble Deep Field the angular correlation
function of galaxies in the redshift interval $[1,1.2]$, which
corresponds to the most likely source redshift for weak lensing surveys. They
found an angular correlation length of 3 arcsec, and therefore negligible
for our purpose. In practice the redshift slice of the lensed galaxies is
much larger $[0.5,1.2]$, which makes the contribution of the angular clustering totally
negligible (Le F\`evre et al. 1996).
As a consequence, the ensemble average of the local variance of the
convergence is simply given by:

\begin{equation}
{\rm P}\left[{\rm A}\left[N^2(\thetag)\right]\right]=
{\sigma_\epsilon^2\over 2 n_g}\int{\rm d}\kg|\tilde W(\kg)|^2,
\label{var_local}
\end{equation}
as derived in Kaiser \& Squires 1993. The noise correlation function is
then

\begin{equation}
{\rm P}\left[{\rm A}\left[N(\thetag)N(\thetag')\right]\right]=
{\sigma_\epsilon^2\over 2 n_g}\int{\rm d}\kg~
e^{i\kg\cdot(\thetag'-\thetag)} |\tilde W(\kg)|^2.
\label{noise_corr}
\end{equation}

\subsection{Average mass error bar}

The average mass error bar $\Delta M$ in an aperture (the total mass of
a cluster for instance, or the mass of a substructure) is now calculated.
Provided that the redshifts of the background galaxies are all the same, the
projected mass $M(\thetag)$ in the direction $\thetag$ is $M(\thetag)
=\Sigma_{crit}(z_l,z_s)K(\thetag)$, where $\Sigma_{crit}(z_l,z_s)$ is the critical surface
mass density of the lens which depends on the lens configuration
(Eq.(\ref{sigma_crit})). If
the background galaxies were distributed in redshift,
the reduced shear becomes redshift dependent but it can be easily
accounted for (Seitz \& Schneider 1997), and with no loss of generality
the analysis is restricted to a single source redshift.

The variance of the total projected mass estimated within an area
$\Sc$ is by definition

\begin{equation}
\Delta M^2=\Sigma_{crit}^2\int_\Sc\int_\Sc~{\rm d}\thetag~{\rm d}\thetag'
{\rm P}\left[{\rm A}\left[N(\thetag)N(\thetag')\right]\right].
\label{var_global}
\end{equation}
According to Eq.(\ref{noise_corr}) the variance is finally

\begin{equation}
\Delta M^2=\Sigma^2_{crit}{\sigma_\epsilon^2\over 2}\int_\Sc{{\rm d}\thetag\over n_g}
\int_\Sc{\rm d}\rg\int{\rm d}\kg~e^{i\kg\cdot\rg}|\tilde W(\kg)|^2.
\end{equation}

\subsection{Gaussian smoothing}

The Gaussian window has a width $\sigma$, and is normalized:

\begin{equation}
W(|\thetag|)={1\over \pi\sigma^2}\exp\left(-{|\thetag|^2\over \sigma^2}\right).
\label{gauss_window}
\end{equation}
Inserting this into Eq.(\ref{noise_corr}) gives the noise correlation function

\begin{equation}
\langle N(\thetag)N(\thetag')\rangle={\sigma_\epsilon^2\over 2}{1\over
2\pi\sigma^2n_g}\exp\left(-{|\thetag'-\thetag|^2\over 2\sigma^2}\right).
\label{noise_corr_gauss}
\end{equation}
The variance defined by Eq.(\ref{var_global}) for a square
area $\Sc=[2a,2a]$ is then straightforward to calculate:

\begin{eqnarray}
\Delta M^2&=&4\sigma^4\left({\sigma_\epsilon^2\over 4\pi\sigma^2 n_g}\right)
\Sigma_{crit}^2\times\nonumber\\
&&\left[
1-\exp\left(-2{a^2\over\sigma^2}\right)-\sqrt{2\pi}{a\over \sigma}
{\rm erf}\left(\sqrt{2}{a\over \sigma}\right)\right]^2.
\label{var_calc}
\end{eqnarray}

\subsection{Top-hat smoothing}

The normalised top-hat of radius $\sigma$ is

\begin{eqnarray}
W(|\thetag|)&=&{1\over \pi \sigma^2} \ \ {\rm if} |\thetag|<\sigma\nonumber\\
&=&0 \ \ {\rm otherwise}
\end{eqnarray}
The noise correlation function is then

\begin{eqnarray}
\langle N(\thetag)N(\thetag')\rangle&=&{\sigma_\epsilon^2\over 2} {1\over
\pi\sigma^2n_g}{s\left(|\thetag'-\thetag|\right)\over \pi\sigma^2}
\ \ {\rm if} |\thetag'-\thetag|<2\sigma\nonumber\\
&=&0 \ \ {\rm otherwise}
\label{noise_corr_tophat}
\end{eqnarray}
where $s\left(|\thetag'-\thetag|\right)$ is the overlapping area
of two identical circles of radius $\sigma$ distant by
${\rm d}=|\thetag'-\thetag|$

\begin{equation}
s({\rm d})=\sigma^2\left[2{\rm arccos}\left({{\rm d}\over 2\sigma}\right)-{{\rm d}\over
\sigma}\sqrt{1-{{\rm d}^2\over2\sigma^2}}\right],
\end{equation}

\section{Cluster lens simulations}
\subsection{Description}

The description of the noise given above is now compared to mass reconstruction simulations using the lens
comparable in lensing strength to a non-critical cluster of galaxies ($\kappa_{max}=0.3$).
The lens potential $\phi$ is
the sum of two non-singular isothermal spheres with a core radius $R_c$, located
at $\thetag_1$ and $\thetag_2$:

\begin{eqnarray}
\psi(\thetag)&=&{4\pi\sigma_\infty^2\over c^2}{\Dc_{ol}\Dc_{ls}\over \Dc_{os}}
R_c^2\Big(\sqrt{1+{|\thetag-\thetag_1|^2\over R_c^2}}\nonumber\\
&+& 0.8\sqrt{1+{|\thetag-\thetag_2|^2\over R_c^2}}\Big).
\label{lens_car}
\end{eqnarray}

The velocity dispersion is $\sigma_\infty=900~{\rm km/s}$,
$R_c=5$, and we assume a lens redshift $z_l=0.2$. The distances
are calculated in an Einstein-de-Sitter Universe.
\begin{figure}
\epsfysize=2.4in
\epsffile{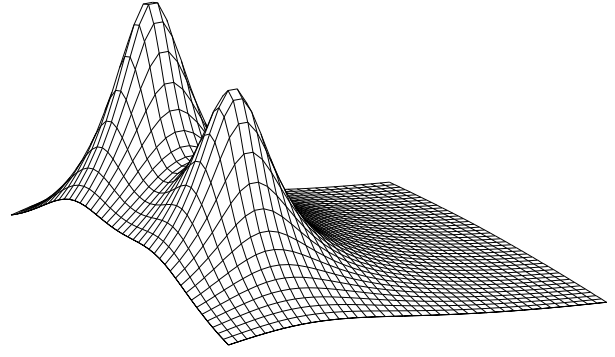}
\caption{\label{massmodel.ps} The cluster mass model used in this work
sampled on a $42\times 42$ pixels grid, with a maximum convergence 
$\kappa_{max}=0.3$.
The lens, defined by Eq.(\ref{lens_car}) has a redshift
$z=0.2$, and an Einstein-de-Sitter Universe is assumed.
This is the mass distribution caused by two truncated isothermal potential
with a core radius of $5$ pixels.}
\end{figure}
Figure \ref{massmodel.ps} shows the convergence $\kappa=\Delta\psi$ of
the lens. The shear caused by the lens is applied to a background galaxy
population at redshift $z_s=1$, with an ellipticity distribution given by

\begin{equation}
p(\epsilon^{(S)})\propto \exp\left[-\left({\epsilon^{(S)}/
\sigma_\epsilon}\right)^2\right],
\end{equation}
with $\sigma_\epsilon=0.12$ (corresponding to a typical axis ratio of $0.8$).
$1000$ background galaxies are randomly placed on a $42\times42$ pixels
grid. An {\it observed} simulated shear map is then calculated, and the lens
mass distribution is reconstructed with a maximum-likelihood algorithm
(Bartelmann et al. 1996).
The maximum likelihood reconstruction minimizes the $\chi^2$ function 
defined as

\begin{equation}
\chi^2={\displaystyle\sum_{i,j}}\left|\ggr(\phi_{guess})-\ggr_{obs}\right|^2,
\end{equation}
\begin{figure}
\epsfysize=3.0in
\epsffile{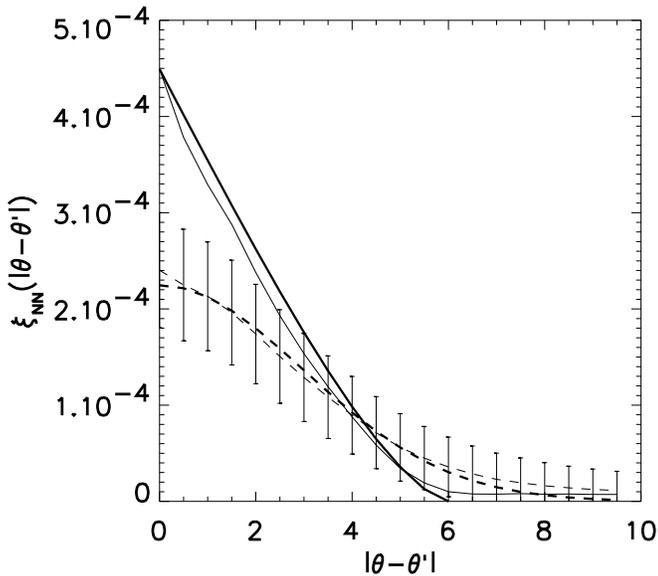}
\caption{\label{correl_bis.ps} The noise correlation function
measured in the
simulations (thin lines), compared to the model (thick lines). The dashed lines
correspond to the Gaussian smoothing and the solid lines for top-hat
smoothing.
Error bars are calculated from 50 different simulation realizations, they are
displayed for the gaussian smoothing only, but they are similar for the
top-hat case.}
\end{figure}
\begin{figure}
\epsfysize=4.4in
\epsffile{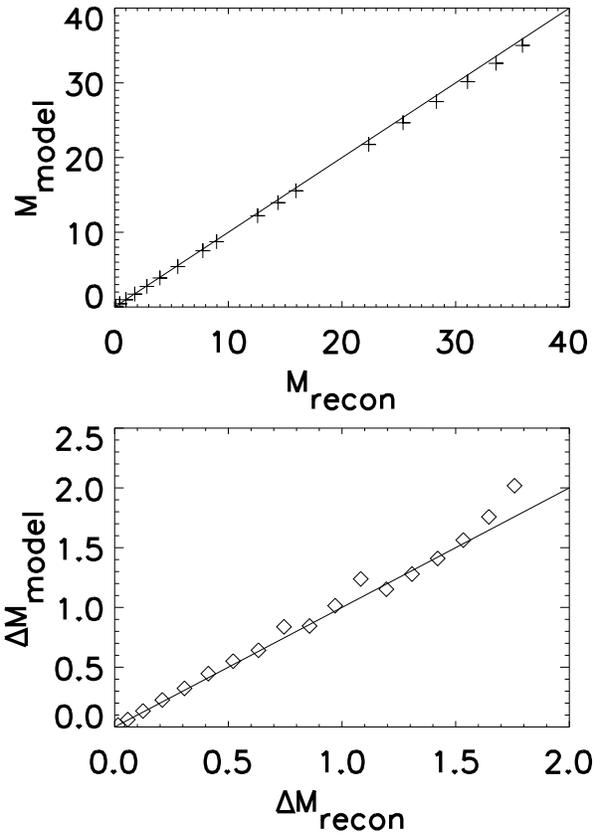}
\caption{\label{tot_error_bis.ps} The crosses on the first plot show the
total mass $M_{recon}$ measured in the simulation for different sizes
of the aperture $S$ and for the gaussian smoothing case. For the smallest aperture,
which corresponds to the
pixel size, $M_{recon}\simeq 0.1$. The largest aperture has a width of
$18$ pixels (about half of the simulation grid), and $M_{recon}\simeq 36$.
The vertical axis displays the smoothed mass distribution of the
model. The solid line shows a perfect correlation between $M_{recon}$
and $M_{model}$. The reconstructed mass is overestimated by less than
$2\%$. The diamonds on the second plot show the standard deviation of
the mass for different aperture sizes. The horizontal axis corresponds to the
measure of $\Delta M$ in the simulation, and the vertical axis is the
prediction from Eq.(\ref{var_calc}). The solid line shows a perfect correlation
between $\Delta M_{recon}$ and $\Delta M_{model}$.}
\end{figure}
where $\phi_{guess}$ is a guessed lensing potential. The reduced shear
$\ggr(\phi_{guess})$ is calculated from $\phi_{guess}$, and $\ggr_{obs}$
is the observed reduced shear. The reduced shear is calculated from the
finite difference scheme described in Van Waerbeke et al. (1999).
The comparisons are done between
the reconstructed mass map and the initial mass smoothed
with the same window used to smooth the shear for the reconstructions (mass
reconstruction and smoothing should commute according to the noise model).
For each type of smoothing window (top-hat and Gaussian) 50 simulations
are performed. The widths of the smoothing windows are
$\sigma=3$ pixels, therefore there is a mean number of $16$
galaxies per aperture. An edge of width $3$ pixels is
systematically removed from the analysis because of its intrinsic
higher noise due to the change of the finite difference scheme at the boundaries of
the grid and the smaller number of galaxies in the aperture.
The calibration constant $\lambda$ (see Section 4.2)
is calculated
by forcing the total mass of each reconstructed map to be equal to the
total mass of the smoothed mass distribution of the lens. Therefore
the variance of the {\it total} reconstructed mass map is zero, so the analysis
is restricted to half of the grid size.

\subsection{Effect of the mass-sheet degeneracy}

For each reconstruction, the mass map is rescaled in such a way that
the total mass measured on each realization is the same as the total
mass in the smoothed lens model.

Such a rescaling changes the amplitude of the noise correlation
function:
if $K_{real}$ is the real mass distribution, the reconstructed mass map
$K_{recon}$ is related to the real mass map by
$K_{recon}=1-\lambda+\lambda K_{real}+N$, where $\lambda$ is the mass-sheet
degeneracy constant and $N$ the noise term. Therefore the estimate
of the real mass distribution is $K_{real}+{N\over \lambda}$ where
the noise scales now with $\lambda^{-1}$.
It follows that the noise correlation function of the reconstructed mass map
scales as $\lambda^{-2}$, according to Eq.(\ref{noise_corr}).
In that sense, error bars of $K$ depend on the absolute value of $K$, although
it is not explicit in Eq.(\ref{noise_corr}), and this rescaling has to be taken
into account when the amplitude of the noise correlation function is calculated.

\subsection{Correlation functions}

Figure \ref{correl_bis.ps} is a comparison of the measured noise correlation
function to the predictions (Eq.(\ref{noise_corr_gauss}) for the Gaussian
smoothing and Eq.(\ref{noise_corr_tophat}) for the top-hat smoothing).
The dashed lines correspond to the Gaussian smoothing and the solid
lines to the top-hat.
The thin lines display the measurement in the simulations, they are in very good
agreement with the predictions (thick lines). As expected, the correlation drops to zero
for the top-hat window when $|\rg|=2\sigma$. The amplitude of the correlation function
is twice the value for the top-hat compared to the Gaussian,
as expected from Eq.(\ref{noise_corr_gauss}) and (\ref{noise_corr_tophat}).

\subsection{Mass error bar}

The average variance of the mass in an area $\Sc=2a\times 2a$ is
given by Eq.(\ref{var_global}). This quantity does not depend on the signal
according to our hypothesis. As shown by the plot at the top of Figure
\ref{tot_error_bis.ps} the reconstructed mass is an unbiased
estimate of the smoothed initial mass distribution (as expected from
Eq.(\ref{itissmooth})). The reconstructed mass is slightly overestimated
by less than $2\%$.

The diamonds on the bottom plot of Figure \ref{tot_error_bis.ps} shows the
error Eq.(\ref{var_calc}) ($\Delta M_{model}$), compared to the measured error on the simulations
($\Delta M_{recon}$).
The different diamonds correspond to different sizes of the smoothing area
$\Sc=2a\times 2a$ ($\Delta_{recon}\simeq 0.1$ for $2a=1~pixel$, and
$\Delta_{recon}\simeq 1.5$ for $2a=18~pixels$). The agreement between the model and the
measurement in the simulation is good, however, there is a tendency of the measured
variance to be slightly underestimated for $2a\simeq 18$. This is a consequence of the
normalisation
of the mass maps (the total measured mass is explicitely set to to the total mass in the
model, so the measured variance is zero at the scale $2a=42~pixels$). The errors are
independent of the position of $\Sc$ on the grid, which confirms that signal
and noise are uncorrelated to a very good approximation.

\section{Significance of peaks}

\begin{figure}
\epsfysize=3.0in
\epsffile{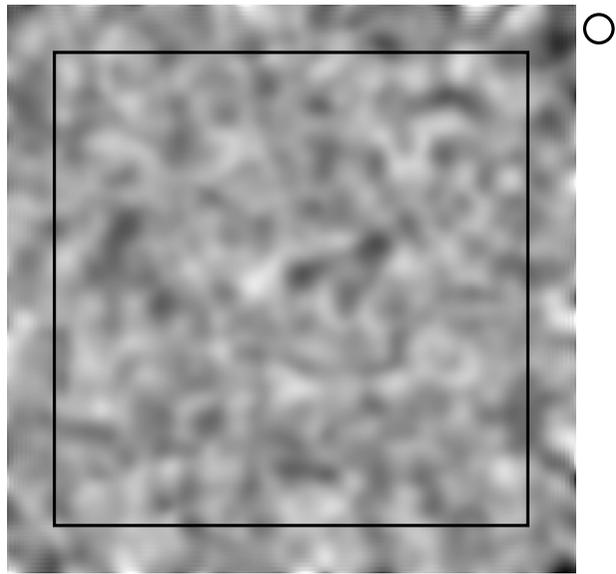}
\caption{\label{recon.ps} One realization of mass reconstruction with no lensing
inside. All the structures here are noise. The high density peak at the center
is a $3\sigma$ fluctuation ($\kappa=0.081$). The dark square represents the subfield
where the statistical analysis is done (in order to avoid the noisier
boundaries), and
the circle at the top right has a $0.66'$ diameter, representing the size of
the Gaussian window width. The field
is $14'\times 14'$ large, with a number density of $30$ galaxies per $arcmin^2$. The intrinsic
ellipticity dispersion is $\sigma_\epsilon=0.17$.}
\end{figure}

The reconstructed mass maps are generally very noisy, even with an arcmin scale
smoothing, and it is difficult to disentangle real and noise over-densities.
Figure \ref{recon.ps} is an example of mass reconstruction
when no lenses are present. It shows fluctuations which can be misinterpreted
as real mass over-densities. It makes particularily difficult the selection of objects
based on their mass, and in practice we use a high signal-to-noise cut-off
to be sure to minimize the number of spurious peak detections
(Schneider 1996, Reblinsky \& Bartelmann 1999).
According to Section 3.2, the noise in the reconstructed mass maps
is equivalent to the convolution of a point process by a smoothing aperture,
therefore it is expected to approach a Gaussian field distribution when the aperture
is larger than the mean angular distance between galaxies.
The nice feature of Gaussian fields is that they are fairly well understood
(Bardeen et al. 1986), in particular the statistics of two-dimensional
Gaussian random fields as been worked out in detail for the case of
cosmic microwave background fluctuations
(Bond \& Efstathiou 1987, hereafter BE).

I propose here to use the statistics of 2D Gaussian random fields as a tool
to analyze the weak lensing mass maps. With this formalism, the significance and
the morphology of the structures can be evaluated analytically and compared wih the
data, which is a way to go
beyond the detection of dark halos done in Schneider (1996), based on a local
signal-to-noise analysis.

\subsection{Height of peaks}

Let $N(\thetag)$, the noise part of Eq.(\ref{kappanoise}), be a 2D
Gaussian random field. The field and its derivatives ($\Vg=[N,N',N'']$)
are distributed as

\begin{equation}
\Pc(\Vg)={1\over (2\pi)^{3/2}|\Mg |^{1/2}} \exp\left(-{1\over 2}
\Vg^t \cdot\Mg^{-1}\cdot\Vg\right),
\label{gauss_proba}
\end{equation}
where the correlation matrix $\Mg$ is

\begin{equation}
\Mg=\left(\matrix{\sigma_0^2 & 0 & -\sigma_1^2 \cr 0 & \sigma_1^2 & 0 \cr -\sigma_1^2 & 0 & \sigma_2^2}\right).
\end{equation}
\begin{figure*}
\epsfysize=4.5in
\epsffile{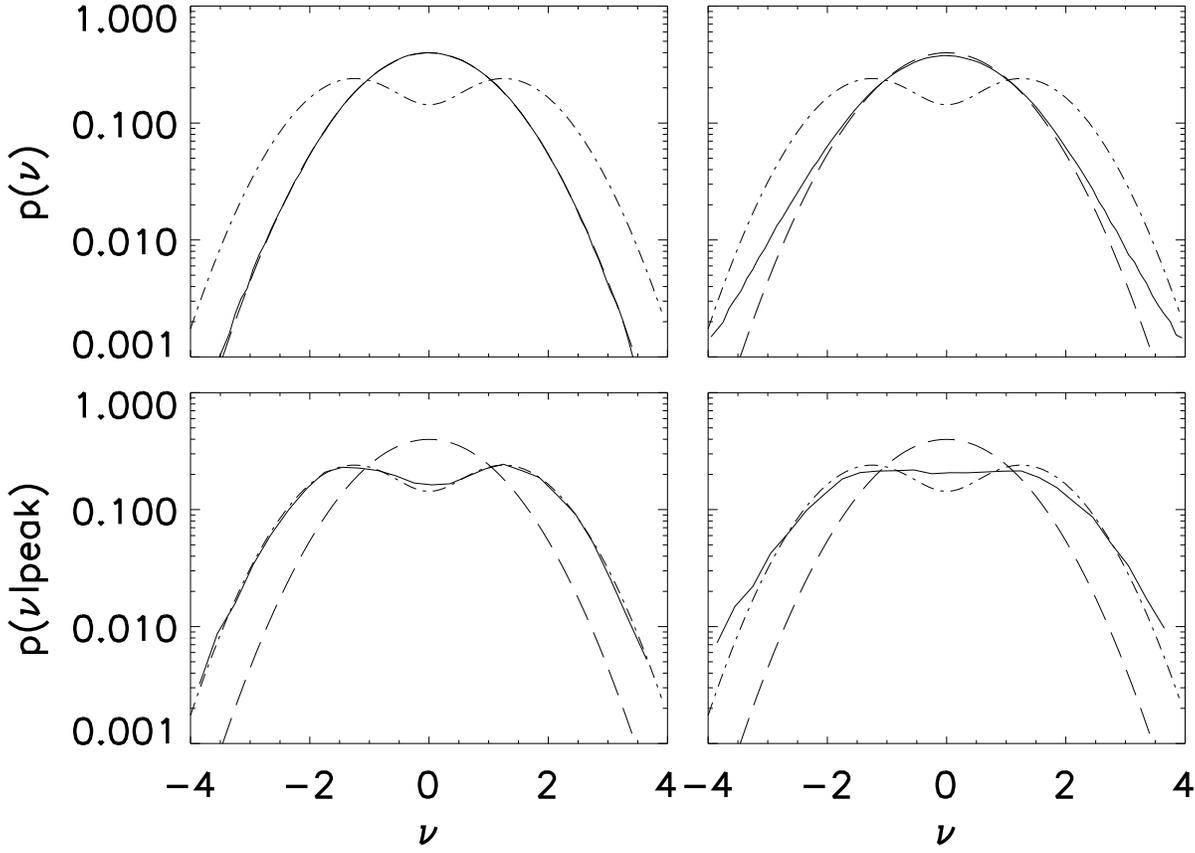}
\caption{\label{field.ps} In all the plots, the dashed curve is
the probability distribution function that a randomly chosen pixel has a
value $\nu\sigma_0$ (Eq.(\ref{field_proba})), and the dot-dashed line is
the expected probability distribution of peaks extrema, according to
Eq.(\ref{peak_proba}). The solid line in the top-left plot is the measured
distribution of all the pixels in the 90 simulations (described in
Section 5.1.)
limited to the sub-field defined by the dark square as explained in Figure
\ref{recon.ps}. The solid line in the bottom-left plot shows the distribution
of extremas measured in the same sub-fields of the simulations. In the
top-right plot, the solid line is the measured distribution of the value
of all the pixels in the whole field, and in the bottom-right
plot, the solid line is distribution of extrema measured in the whole field.
}
\end{figure*}
The $\sigma_n$ are determined from the noise power spectrum

\begin{equation}
\sigma_n^2=\int {k{\rm d}k\over 2\pi} k^{2n} \langle |\tilde N(k)|^2\rangle,
\end{equation}
which can be calculated from the two-point correlation function:

\begin{equation}
\sigma_n^2=(-1)^n{2\pi\over \int_0^{2\pi}{\rm d}\phi \cos^{2n}(\phi)}
\left({{\rm d}^{2n}\langle N(\thetag)N(\thetag+\rg)\rangle\over {\rm d}r^{2n}}\right)_{r=0}.
\end{equation}
The zero-lag value of the correlation function is $\sigma_0^2=\langle N(0)N(0)\rangle$.
The marginalisation of Eq.(\ref{gauss_proba}) over the field derivatives
gives the -Gaussian- probability that any pixel randomly chosen in the
field has the value $\nu\sigma_0$:

\begin{equation}
\Pc(\nu)={1\over \sqrt{2\pi}} \exp \left( -{\nu^2\over 2}\right).
\label{field_proba}
\end{equation}
In order to describe the probability distribution function of peaks, the
following parameters are introduced:
$\gamma$ and $\theta_\star$ are respectively
the width of the power spectrum and the characteristic angular scale
of the field:

\begin{eqnarray}
\gamma={\sigma_1^2\over \sigma_0\sigma_2}; \ \ 
\theta_\star=\sqrt{2}{\sigma_1\over \sigma_2}.
\end{eqnarray}
A peak occurs, by definition, where the gradient of the field vanishes
($\nabla_\thetag N=0$), and it has a given height $\nu\sigma_0$. Its
probability distribution function $\Pc(\nu |peak)$ can be calculated
from the number density of minima and maxima $\Nc_{max}(\nu)$ and $\Nc_{min}(\nu)$
(see BE for the details):

\begin{equation}
\Pc(\nu|peak)={\Nc_{max}(\nu)+\Nc_{min}(\nu)\over \int_{-\infty}^\infty{\rm d}\nu
(\Nc_{max}(\nu)+\Nc_{min}(\nu))},
\label{peak_proba}
\end{equation}
where

\begin{equation}
\Nc_{max}(\nu)=\Nc_{min}(-\nu)={1\over 2\pi\theta_\star}\exp
\left(-\nu^2/2\right){G(\gamma,\gamma\nu)\over \sqrt{2\pi}}.
\end{equation}
The function $G(\gamma,\bar x)$ depends on the variable $b^2=2(1-\gamma^2)$:

\begin{eqnarray}
G(\gamma,\bar x)&=&{1\over 2}\left(\bar x^2+{b^2\over 2}-1\right){\rm erfc}
\left(-{\bar x\over b}\right)+{\bar x b\over 2\sqrt{\pi}}e^{-{\bar x^2\over b^2}}
\nonumber\\
&+&{1\over 2(1+b^2)^{1/2}} e^{-{\bar x^2\over 1+b^2}}{\rm erfc}\left(-{\bar x\over
b\sqrt{1+b^2}}\right).
\label{G_func}
\end{eqnarray}
The denominator of Eq.(\ref{peak_proba}) has an analytical solution which can be derived
straightforwardly with a formal calculator, but it is not worth to give its expression here.

For a Gaussian smoothing (Eq.(\ref{gauss_window})) the spectral parameters take the
values $\gamma=1/\sqrt{2}$ and $\theta_\star=\sigma/\sqrt{2}$, and therefore
$\Pc(\nu|peak)$ only depends on $\nu$. The denominator of Eq.(\ref{peak_proba}) is:

\begin{equation}
\int_{-\infty}^\infty{\rm d}\nu(\Nc_{max}(\nu)+\Nc_{min}(\nu))={1\over (2\pi)^{3/2}\theta_\star}{4\sqrt{3}\pi-9\over 9\sqrt{2\pi}}
\end{equation}
It means that once the mass map is
normalised by $\sigma_0$ the probability distribution of peaks is always the same,
independently of the smoothing scale, the number density of background galaxies, and the ellipticity
dispersion $\sigma_\epsilon^2$.

The analytical predictions of Eq.(\ref{field_proba}) and Eq.(\ref{peak_proba}) are
now compared to numerical simulations. Ninety mass reconstructions have been performed,
with no lensing signal inside. The total field of view for each realization is
$4100\times 4100$ pixels (equivalent to a $14'\times 14'$ area at the Canada France Hawaii
Telescope), and it contains about
$30~gal/arcmin^2$. The intrinsic ellipticity dispersion is $\sigma_\epsilon=0.17$ (using the distribution
Eq.(\ref{elli_distrib})), and the
smoothing aperture is given by Eq.(\ref{gauss_window}) with $\sigma=0.33'$, which corresponds
to an average of 10 galaxies per aperture. The "mass" field is reconstructed using the
maximum-likelihood algorithm. Since the aperture is truncated at the edge of the field, only a sub-set of pixels in the simulations are used, those
enclosed in the square area defined on Figure
\ref{recon.ps}. This Figure also shows the scale of the smoothing window compared
to the whole field.

%

Figure \ref{field.ps} shows the measured probability distribution
functions in the field and for peaks, for the whole field of the
simulations, and for the sub-field as described in Figure \ref{recon.ps}.
The measurements are represented by the solid lines, while the dashed
and dot-dashed curves show respectively the unconstrained
(Eq.(\ref{field_proba})) and the constrained (Eq.(\ref{peak_proba}))
distributions.
For the centerred sub-field it is remarkable that the peak distribution prediction fits so well the
measurements. It shows that the non-linear ML reconstruction give a noise
comparable to the noise in weak lensing approximation, confirming our previous results. However it
is clear that when the measurement is done in the
whole field, the statistic is not gaussian. This is an effect caused by the boundary
of the reconstruction: at the edge of the field, the smoothing
aperture is truncated and it contains less galaxies than in the central part
of the simulations. Therefore the peak analysis has to be restricted
to the area where the aperture is not too close to the edge.
Two other sets of 20 simulations each, one with $\sigma_\epsilon=0.4$ and one with a
uniform distribution of ellipticities between -0.2 and 0.2 (instead of Eq.(\ref{elli_distrib}))
give the same result, which shows that the Gaussian behavior of the noise does not depend on the
details of the ellipticity distribution. The distribution of peaks is not Gaussian, as expected
for Gaussian fields. It shows that simple signal-to-noise analysis based on the local
variance with Gaussian probabilities is wrong. For instance a value at $3\sigma$
in the field, with $\Pc(\nu)=0.004$
corresponds to $\Pc(\nu|peak)=0.03$ for a peak, almost a factor of ten higher.

Because of the discreteness of the spatial distribution of the galaxies a
Poisson-like noise is expected rather than a Gaussian noise.
The fact that the noise is not Poisson is a consequence of the central
limit theorem which makes the distribution more Gaussian because there is
an average of $10$ galaxies per aperture, which is much larger than $1$. 

The peak statistic described here is very easy to implement and it
requires only the calculation of $\sigma_0$. In other words, only the galaxy number
density, the variance of their ellipticities and the shape of the smoothing window
are needed.

\subsection{Profile of peaks}

\begin{figure}
\epsfysize=2.7in
\epsffile{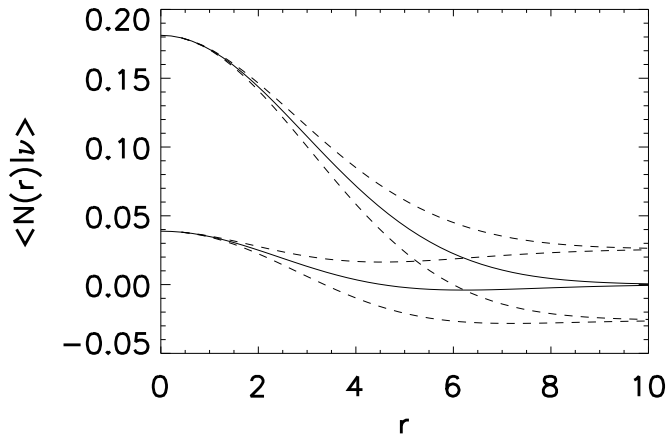}
\caption{\label{profile_new.ps} Mean profile around a peak of height $7\sigma_0$
(top curve) and $1.5\sigma_0$ (bottom curve) with the noise correlation
function given by Eq.(\ref{noise_corr_gauss}). The parameters are
$\sigma=3$, $n_g=0.6~{\rm galaxies/pixel^2}$, and $\sigma_\epsilon=0.12$.
The dashed curves represent the $\pm~1\sigma$ deviations.}
\end{figure}

It is useful to add the morphological information to the
peak analysis in order to constrain the mass distribution better.
Let $\Pc(shape|\nu)$ be the probability that a peak of
height $\nu\sigma_0$ has a given shape, and $\Pc(shape,\nu)$
the global peak probability written as:

\begin{equation}
\Pc(shape,\nu)=\Pc(\nu|peak)\Pc(shape|\nu)
\label{all_proba}
\end{equation}
BE give the analytical formulae describing the mean profile of a peak and
its dispersion. This calculation depends on the averaged curvature
of a peak of height $\nu$. The curvature $x$ is defined from the
eigenvalues $(\lambda_1,\lambda_2)$ of the second derivative
matrix of the noise $\partial_{ij}N$:

\begin{equation}
x=(\lambda_1+\lambda_2)/\sigma_2.
\end{equation}
Following BE, all distances and derivatives are calculated with respect to
the reduced coordinates $\omegag=\rg/\theta_\star$. The reduced correlation
function is defined as $\psi(\omegag)=\langle N(\thetag)N(\thetag+\omegag)
\rangle/\langle N(0)N(0)\rangle$.
The profile around a peak, averaged over the curvature and the ellipticity,
is Gaussian distributed with a mean
$\langle N(\omegag)|\nu\rangle$ and a dispersion
$\langle N^2(\omegag)|\nu\rangle$ (Eqs. A2.9 and A2.8 of BE):

\begin{eqnarray}
\langle N(\omegag)|\nu\rangle/\sigma_0&=&{\nu\over 1-\gamma^2}(
\psi+\Delta\psi/2)-{\langle x|\nu\rangle/\gamma\over 1-\gamma^2}(\gamma^2\psi+\Delta\psi/2)
\nonumber\\
\langle N^2(\omegag)|\nu\rangle/\sigma_0^2&=&1-{1\over 1-\gamma^2}\psi^2
-{1\over 1-\gamma^2}\psi\Delta\psi\nonumber\\
&-&{1\over \gamma^2(1-\gamma^2)}(\Delta\psi/
2)^2-{1\over \gamma^2}(\psi')^2\nonumber\\
&-&{1\over 2\gamma^2} \left[\omega (\psi'/\omega)'\right]^2.
\label{profiles}
\end{eqnarray}
For 2D Gaussian fields, the conditional mean value of the average curvature
$\langle x|\nu\rangle$ can be calculated analytically:

\begin{equation}
\langle x|\nu\rangle=\gamma\nu+{H(\gamma,\gamma\nu)\over G(\gamma,\gamma\nu)}.
\end{equation}
The function $H(\gamma,\bar x)/G(\gamma,\bar x)$ tends to zero for high peaks
($\nu\rightarrow \infty$). $G(\gamma,\bar x)$ was given in Eq.(\ref{G_func})
and $H(\gamma,\bar x)$ is:

\begin{eqnarray}
H(\gamma,\bar x)&=&
{\bar x b^2\over 2}{\rm erfc}\left(-{\bar x\over b}\right)
+{b^5\over 2\sqrt{\pi}(1+b^2)}e^{-{\bar x^2\over b^2}}\nonumber\\
&-&{\bar x b^2\over 2(1+b^2)^{3/2}} e^{-{\bar x^2\over 1+b^2}} {\rm erfc}
\left(-{\bar x\over b\sqrt{1+b^2}}\right)
\end{eqnarray}
The quantity $\chi^2$ quantifies how any observed peak profile
of height $\nu\sigma_0$ $N_{obs}(\thetag,\nu)$ is different from
its expectation value:

\begin{equation}
\chi^2={\displaystyle \sum_\omegag}{\left(N_{obs}(\omegag,\nu)-
\langle N(\omegag)|\nu\rangle\right)^2\over \langle N^2(\omegag)|\nu\rangle}.
\end{equation}
This follows a $\chi^2$ statistic with the number of degrees of freedom $n$
given by the number of positions $\omegag$ involved in the summation. Thus the
probability $\Pc(shape|\nu)$ that a peak of height $\nu\sigma_0$ having
a definite shape is a noise fluctuation is,

\begin{equation}
\Pc(shape|\nu)={1\over 2\Gamma\left({n\over 2}\right)}\int_{\chi^2}^\infty \left(
{t\over 2}\right)^{{n\over 2}-1} e^{-{t\over 2}}.
\label{shape_proba}
\end{equation}
The probability $\Pc(shape,\nu)$ that a peak of height $\nu\sigma_0$ with a
given shape is a noise fluctuation is then given by Eq.(\ref{all_proba}),
which contains the probabilities (\ref{peak_proba}) and (\ref{shape_proba}).
Figure \ref{profile_new.ps} is an example of mean profile and its dispersion
derived from Eqs.(\ref{profiles}).

\section{Conclusion}

It is possible to understand the noise properties
of reconstructed mass maps with rather simple arguments.
The noise in ML mass reconstruction is very close
to a 2D Gaussian random field, which has several implications for the analysis
of mass maps: for cluster lensing,
the "blurred" area around cluster of galaxies where
the identification of mass clumps and substructures is uncertain can
in fact be described accurately. It is possible to give the probability that
the fluctuations are real mass-over-densities, and to give their mass
with an error bar. For large scale surveys, small mass concentrations
(not visible individually) can be detected statistically as a deviation from
the expected noise Gaussian statistic. Groups and
small clusters should show up in the statistical distribution
of peaks. Preliminary results using ray-tracing simulations (Jain, Seljak,
White 1999, Jain private communication) already confirmed this, but the
detailled analysis of peak statistic in realistic cosmological models
is left for another work.

The noise autocorrelation function and the mass error bars were
derived and compared to numerical simulations of non-critical
cluster lenses. The agreement between the predictions
and the simulations proves that the noise, caused by the intrinsic ellipticities
and the discrete distribution of the galaxies, is nearly uncorrelated with the
lensing signal despite the non-linear lensing equation. Consequently, the weak
lensing approximation applies, which simplifies considerably the noise analysis.
It is worth to remind that this is true for non-critical clusters only. However
it is reasonable to think that the
smearing of the signal due to the smoothing implies that the
approximation still works for noise analysis in critical clusters (if the smoothed mass
distribution becomes non-critical).

A BBKS-type approach was used to predict the statistics of noise peaks in mass maps.
Again the numerical simulations are accurately fitted by these predictions. The fact
that the noise approaches a Gaussian distribution is due to the large mean number of galaxies
in each aperture (about $10$). There is certainly
a limitation of the method used here: if the smoothing window is too small or too "fuzzy" it
should deviate from a Gaussian statistic. This departure from Gaussianity was observed
in the S-statistic of Schneider (1996) because he used a compensated filter.
A work worth to be done is to investigate the limits of the method, and eventually to
consider the Poisson behavior of counts in small cells, as in
Szapudi \& Colombi (1996). However, in the context discussed here, it is reliable to use Gaussian
statistics.

I discussed so far only the ML mass reconstruction method, and compared it
to the
analytical KS. Can this analysis be used for
other mass reconstruction methods? There are some numerical indications
that the curl-free line-integral approaches are close to the KS reconstruction,
(Seitz \& Schneider 1996, Lombardi \& Bertin, 1998b)
and therefore close to the ML as well. For the newly
developed regularization methods (Seitz et al. 1998, Bridle et al.
1998) it probably not the case. The noise is reduced in these methods,
in an optimal way that real structures should still emerge from the noise.
The main advantage of the regularization methods is
their adaptative spatial resolution depending on the local signal-to-noise. However the
regularization parameter is arbitrary and can potentially kill the small
scale fluctuations that we are looking for, so the analysis done here should not apply.  
It is therefore valuable to make a study of the noise properties for all the
reconstruction methods in order to search for a better use of the
noisy regions of the mass maps and to compare their ability
to detect the mass overdensities.

\section{acknowledgments}

I thank R. Scoccimarro, Y. Mellier and T. Erben for a carefull reading of
the manuscript.


\begin{thebibliography}{}

\bibitem{} Bardeen, J. M., Bond, J.R., Kaiser, N., Szalay, A.S., 1986, ApJ, 304, 15
\bibitem{} Bartelmann, M., Narayan, R., Seitz, S., Schneider P., 1996, ApJ,
464, L115
\bibitem{} Bond, J.R., Efstathiou, G., 1987, MNRAS, 226, 655 (BE)
\bibitem{} Bridle, S.L., Hobson, M.P., Lasenby, A.N., Saunders, R., 1998, MNRAS,
299, 895
\bibitem{} Connolly, A.J., Szalay, A.S., Brunner, R.J., 1998, ApJ, 499, L125
\bibitem{} Erben, T et al., astro-ph/9907134
\bibitem{} Hattori, M., 1997, "Cosmic Chemical Evolution", IAU 187, Kyoto, Japan
\bibitem{} Jain, B. Seljak, U., White, S., astro-ph/9901191
\bibitem{} Kaiser, N. 1995, ApJ, 439,L1
\bibitem{} Kaiser, N. 1998, ApJ 498, 26
\bibitem{} Kaiser, N., Squires, G., 1993, ApJ 404, 441
\bibitem{} Kaiser, N. et al., astro-ph/9809268
\bibitem{} Le F\`evre, O., Hudon, D., Lilly, S.J., Crampton, D., Hammer, F., Tresse, L., 1996, ApJ, 461, 534
\bibitem{} Lombardi, M., Bertin, G., 1998a, A\& A, 335, 1
\bibitem{} Lombardi, M., Bertin, G., 1998b, A\& A, 330, 791
\bibitem{} Lombardi, M., Bertin, G., 1999, A\& A, 348, 38
\bibitem{} Oukbir, J., Blanchard, A., 1997, A\& A, 317, 1
\bibitem{} Peebles, P.J.E., 1980, {\it The large scale structure of the
Universe}, Princeton
\bibitem{} Reblinsky, K., Bartelmann, M., 1999, A\& A, 345, 1
\bibitem{} Reblinsky, K., Kruse, G., Jain, B., Schneider, P., astro-ph/9907250
\bibitem{} Schramm, T., Kayser, R., 1995, A\& A, 299, 1
\bibitem{} Schneider, P., 1996, MNRAS, 283, 837
\bibitem{} Schneider, P., Seitz, C., 1995, A\& A, 294, 411
\bibitem{} Seitz, S., Schneider, P., 1996, A\& A, 305, 383
\bibitem{} Seitz, C., Schneider, P., 1997, A\& A, 318, 687
\bibitem{} Seitz, S., Schneider, P., Bartelmann, M. 1998, A\& A, 337, 325
\bibitem{} Squires, G., Kaiser N., 1996, ApJ 473, 65
\bibitem{} Szapudi, I.,  Colombi, S., 1996, ApJ, 470, 131
\bibitem{} Van Waerbeke, L., Bernardeau, F., Mellier, Y., 1999, A\& A, 342, 15
\bibitem{} Wilson, G., Cole, S., Frenk, C.S., 1996, MNRAS, 280, 199

\end{thebibliography}
\end{document}